# Hard X-ray observations of the X-ray pulsar A0535+26[a]

**V. Borkous[1], A.S. Kaniovsky[1], R.A. Sunyaev[1], V.V. Efremov[1],**
**P. Kretschmar[2], R. Staubert[2], J. Englhauser[3], W. Pietsch[3]**

[1]Space Res. Inst., Russia Acad.of Sciences, 117810, Profsoyuznaya 84/32, Moscow
[2]Astr. Inst. der Univ. Tuebingen, Waldh. 64, D-72076 Tuebingen, Germany
[3]Max-Planck-Institut fur extraterrestrische Physik, D-85740 Garching, Germany

We present the results of the observations of the "giant" bursts from the X-ray pulsar A0535+26 made by HEXE onboard Mir-Kvant in April 1989, November 1993 and February 1994. The pulse periods were measured, pulse profiles in different energy bands were produced, and their variability was investigated. The power density spectra (PDS) in $2 \times 10^{-3}$-1 Hz range is presented, which shape is typical for flicker-noise processes, usually observed in black hole candidates. The noise rms grows with energy from ~20% at 20 keV to ~30% at 80 keV. The source photon spectrum in the 15-200 keV energy range and its variability over the pulse phase are reported. Approximately the shape of the spectrum can be described by the "canonical" model for X-ray pulsars with power-law index $\gamma$~1.1, cut-off energy $E_c$~23 keV and folding energy $E_f$~19 keV. All these parameters are weakly dependent on the luminosity. The most significant deviation from this continuum is observed at ~100 keV in the spectrum of the main pulse maximum. This feature is interpreted as a cyclotron line. Comparison of the HEXE data with the data from BATSE/CGRO (Bildsten et al., 1997) shows that in the high luminosity state ($L$~$10^{38}$ erg/s) the pulsar's pulse profile differs substantially from the pulse profile in the low-luminosity ($L$~$5 \times 10^{36}$ erg/s) state. This difference is explained by the qualitative change of the polar cap structure with formation of the accretion columns.

## INTRODUCTION

The recurrent pulsar A0535+26 was discovered by the Ariel-5 observatory about twenty years ago (Rosenberg et al., 1975). Soon after that its period of pulsations(Bradt et al., 1976) was found (~103.8 s), and the spectra in different X-ray energy bands were measured (Ricketts et al., 1975; Bradt et al., 1976; Ricker et al., 1976). Almost immediately the pulsar was identified with its optical counterpart: HD 245770 of spectral class O9.7 IIe (Stier, Liller, 1976; Li et al., 1979).

Most of its time the source spends in a quiet state with the X-ray flux 5-10 mCrab, but with a $110^d$ period its luminosity grows substantially: "normal" outbursts are observed. During these outbursts the flux reaches hundreds of mCrabs. This periodicity is not absolute -- sometimes it breaks down, and after the next 110 days the system continues to stay in the quiet state.

Sometimes instead of the normal outbursts (or between them) "giant" outbursts are observed, during which the flux significantly exceeds 1 Crab. Currently four giant outbursts (in May 1975, October 1980, April 1989 and February 1994), about a dozen of normal outbursts and a about dozen of events when a normal outburst was expected but was not happened are known (Motch et al., 1991; Kendziorra et al. 1994; Finger et al. 1996; Chichkov et al., 1997).

Nagase et al. (1982) have assumed that the outbursts are connected with the pulsar motion in the binary system. Using this assumption they made an estimate of the binary period $P_{orb}$. Almost immediately Priedhorsky and Terrell (1983) confirmed the existence of this period using Vela 5B data. Recently Motch et al. (1991) and Finger et al. (1994) studying the normal outbursts received $P_{orb}$=111$^d$.38±0$^d$.11 and $P_{orb}$=110$^d$.3±0$^d$.3, respectively. Finger et al. (1994) reported also the other parameters of the binary system: the periastron passage time $\tau$=JD 2449169.5±0.6, $a_x \sin i$=267±13 lt. s., $\omega_x$=130°±5°, $e$=0.47±0.02. A less accurate estimate of this parameters based on the assumption of the $111^d$ binary period is given by Sembay et al. (1990).

The source broadband spectrum during the giant outburst (Kendziorra et al., 1994) can be roughly described by the "canonical" model for X-ray pulsars (White et al., 1983). Up to ~25 keV this spectrum roughly follows a power law with the photon index $\gamma$~1.1; at the higher energies it drops exponentially as $E^{-\gamma} \exp(-E/E_f)$ with $E_f$~20 keV (see also Dal Fiume et al., 1988; Cusumano et al., 1992; Grove et al., 1995).

The phase resolved spectra differ significantly one from another, and the pulse shape strongly depends on the energy band (Frontera et al., 1985; Dal Fiume et al., 1988; Kendziorra et al., 1994). At ~100 keV the source spectrum

---

significantly deviates from the "canonical" shape. Kendziorra et al. (1994) and Grove et al. (1995) interpreted this feature as a cyclotron absorption line. Kendziorra et al. (1994) also found indications for the line at ~50 keV.

The source power density spectrum (PDS) is roughly described by a two-component model with a flat part at lower (ν<0.02 Hz) frequencies and a power-law drop with index 1-2 at higher frequencies (Frontera et al., 1985; Finger et al., 1996). The PDS deviates from this shape at 20-70 mHz where a peak of a quasi-periodic oscillations is observed (Finger et al., 1996).

OBSERVATIONS

HEXE (Reppin et al., 1983) is a phoswich detector system consisting of four units having a total collecting area of about 800 cm$^2$. The crystals in each detector are 3.2 mm NaI(Tl) backed by 50 mm CsI(Tl). The field of view is restricted by a tungsten collimator to 1.6x1.6 degrees FWHM. The instrument allows to measure the source spectra in the energy range 20-200 keV with 30% resolution at 60 keV. The observations are performed in either collimator rocking mode (2 detectors on source and 2 detectors on background with 2.3 degrees offset) or station rocking mode. The duration of on source and background pointings are 80-180 s each. The low orbit of the Mir station does not allow continuous monitoring of the source and the observations are done in "sessions". The duration of a typical session is between 7 and 30 minutes.

To correct the data for collimator efficiency the exact orientation of the Mir station has to be known. Currently it is determined from simultaneous observations by the TTM telescope also installed onboard Mir-Kvant. The accuracy of the pointing determination is several arc minutes. The calibrations of the third and the fourth detector show that their physical characteristics differ substantially from those of the first pair. In particular they have worse energy resolution. So for source spectroscopy we used the first two detectors only; the third and the fourth detectors were used in timing and pulse profile analysis.

The HEXE telemetry channel allows two modes of observations -- a direct transmission mode (DTM) and a buffered transmission mode (BTM). In both cases the information flow is restricted. In DTM it is impossible to transmit more than 40 counts/s/det, and the data must be corrected on the integral counters readout; in BTM the spectrum is accumulated every 0.85 s, so the time resolution is worse than in DTM. However the count rate from A0535+26 is so high (>100 counts/s/det) that BTM becomes more informative than DTM. Because of the strong telemetry overflow, timing analysis in DTM was performed using two onboard counters accumulating the total number of events registered from the detectors pairs.

Mir-Kvant observed A0535+26 three times: in 1989April, 1993 November and 1994 February (see Table 1). In these observations triggered by IAU circulars announcing outbursts, the flux in the 15-200 keV energy range in all cases was higher than 2.5 Crab. This classifies the outbursts as "giant".

TIMING

The results of the period measurements using the standard epoch folding procedure (Leahy et al., 1983) are presented in Table 1[b]; the pulse profiles are shown in Fig. 1-3.

The pulse profiles of A0535+26 changed substantially during the observations. It is observed both fast variability with a characteristic time scale much shorter than the period of pulsations, and much slower changes with a characteristic time scale greater than the period of pulsations. From the shape of the light curve (Fig. 4) it is possible to suggest that the fast variations are caused by a large number of flares superimposed over the pulsed component.

To analyze this variability we make use of the PDS technique, developed by Leahy et al. (1983). The received PDS were renormalized to remove their explicit dependence on the count rate (see Belloni, Hasinger, 1990). For the time analysis we chose only continuous parts of the light curve (i.e. containing no telemetry losses and collimator switches) with length 41 s.

---

[b] It should be noted that the 1994 February observations were made with large gaps between the adjacent sessions, and the pulse period changed substantially during these gaps. On the other hand, it is impossible to measure such a large period during a single session of the Mir-Kvant observations. So, the reported pulsar period was obtained for JD 2449399 taking into account the pulsar spin acceleration. According to Finger et al. (1996) this acceleration was $\dot{\nu} \approx 1\text{-}1.2\times10^{-11}$ Hz/s at JD 2449399. The epoch-folding technique did not allow also to choose one from the two most probable pulse periods 103.34 s and 103.215 s. However, the value 103.34 s seems to be slightly more preferable.

The typical PDS obtained during the 1994 February observations is presented in Fig. 5. It can be well described by a power law (see Table 2) $P(f_n)(\frac{f_n}{f})^{\gamma}$ with an index $\gamma \sim 1$ (see Table 2); $P(f_n)$ is the normalization factor, $f_n$ is the frequency at which the normalization is made.

The obtained power density spectra has an obvious drawback: the absence of information about the harmonics with frequencies lower than 0.05 Hz. We tried to extend the bandwidth by making the Fourier transform over the whole light curve (including places where telemetry losses and collimator rocking happened). Before making the transform the light curve was corrected for differences in the efficiencies of the detectors. To exclude the input to the PDS from the slow background changes we restricted the frequency range to $2\times10^{-3}$-0.7 Hz.

It is clear that the above procedure is not completely correct. Telemetry losses and collimator rocking can distort the PDS shape. However, this distortions can be minimized if to each bad point one assignes the count rate value from the point preceding it. If the number of bad points is much smaller than the total number of the points in the light curve then the distortion of the PDS will be small. Note, that the correctness of the result may be checked by comparing the high-frequency part of the broadband PDS with the PDS obtained using the standard procedure. The HEXE results are in a good agreement with the observations reported by Frontera et al. (1985).

The power density spectrum, obtained on the basis of the above discussion, is shown in Fig. 6. It is well described by the law $P^{fit}(f) = \begin{cases} P(0), & f < f_{br} \\ P(0)(\dfrac{f}{f_{br}})^{-\gamma}, & f > f_{br} \end{cases}$ ($f_{br}$ is the frequency at which the spectral break happens) using the method described in Borkous et al. (1995). The results of the approximation are presented in Table 2. It follows from these fits that $f_{br} \sim 0.02$ Hz.

A0535+26 is not the only source with a powerlaw-like spectrum. Many black hole candidates have a similar PDS shape. This shape is generated by a flicker noise, i.e. by the multiple X-ray flares. According to Letho (1989) these flares are shorter than $(2\pi f_{br})^{-1} \sim 8$ s. Taking into account the uncertainty of $\nu_{br}$ it is possible to put the $1\sigma$ upper limit $\tau < 20$ s.

The flares are shorter than the pulse period, so the fast aperiodic changes can be explained by the flaring activity. The similar PDS shape is observed in many other X-ray pulsars like Her X-1, Vela X-1, 4U1626-67, Cen X-3, GX 301-2 (Beloni, Hasinger, 1990).

It can be assumed that averaging a large number of the individual pulses will give a result independent of the particular set of the observations. However, the flare activity is very strong (mean root square variation, defined by the formula $rms = \sqrt{\int \tilde{P}(f)df} = \sqrt{\sum_j \tilde{P}_j \Delta f_j}$ reaches 25% of the total count rate) and averaging must be done over a really large data set. For example, the pulse profiles presented in Fig. 2 are produced using time intervals with duration of hundreds of seconds, but the difference in the pulse shapes is still visible.

Note, that from the power density spectra with low time resolution only harmonics with frequencies 1-5 times the pulsation frequency are excluded. This implicitly assumes that the amplitudes of all other harmonics are much smaller than the amplitude of the underlying continuum. To test the validity of this assumption we intentionally left harmonics, the frequencies $\nu$ of which were multiples of the pulsation frequency with $\nu>5/P$, in the high-resolution spectrum, and compared it with the PDS from which these points were removed. No significant discrepancy at $\nu>5/P$ was found, confirming the correctness of the above suggestion.

Table 1. Observation summary

| Data | JD 2440000+ | On-source time, s | Period, s | $L_X^*$, $10^{36}$ $d^2$ ergs/s, 20-100 keV | Orbital phase | HEXE mode |
|------|-------------|-------------------|-----------|---------------------------------------------|--------------|-----------|
| 8/4/89 | 7625 | 2830 | 103.26±0.02 | 5.1±0.3 | 0.09 | DTM |
| 9/4/89 | 7626 | 2530 | 103.26±0.02 | 5.2±0.3 | 0.1 | DTM |
| 14/4/89 | 7631 | 1780 | 103.25±0.02 | 4.2±0.4 | 0.145 | BTM |
| 16/11/93 | 9307 | 427 | 103.4 | 9.5±0.1 | 0.25 | DTM |
| 16/2/94 | 9399 | 450 | 103.34±0.005 | 9.9±0.9 | 0.086 | DTM |
| 17/2/94 | 9400 | 647 | - | 9.7±0.8 | 0.095 | DTM |
| 19/2/94 | 9402 | 600 | - | 9.2±0.8 | 0.112 | DTM |
| 20/2/94 | 9403 | 525 | - | 9.7±0.8 | 0.121 | DTM |

* Assuming an isotropic point source at distance $d$ in kpc.

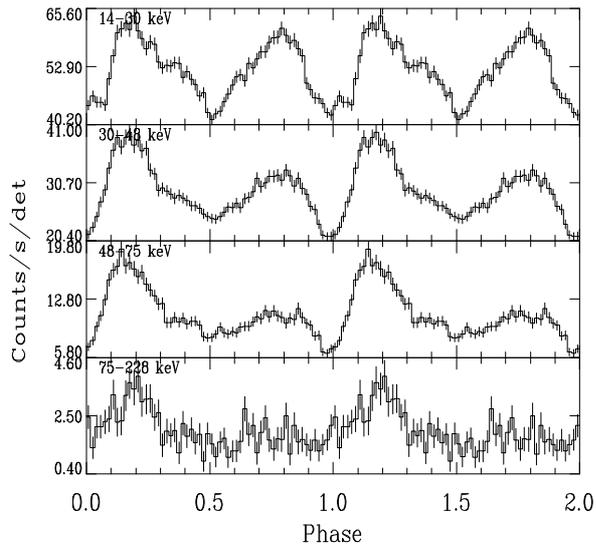

8 February 1989

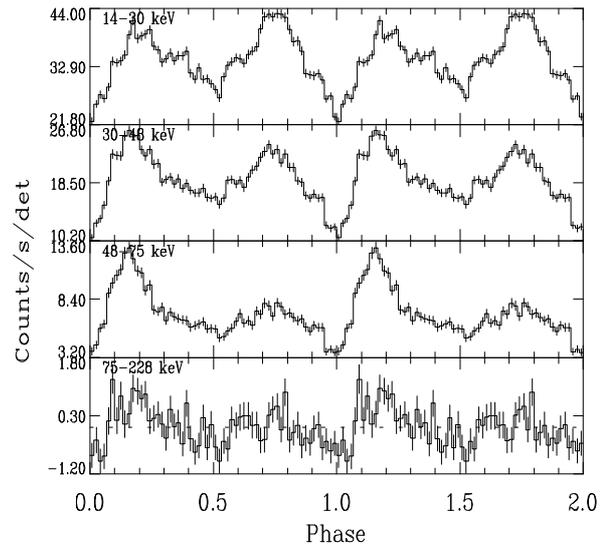

14 February 1989

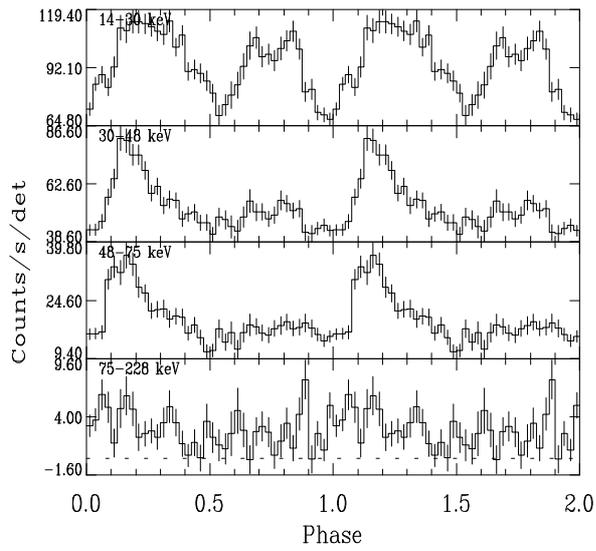

3 November 1993

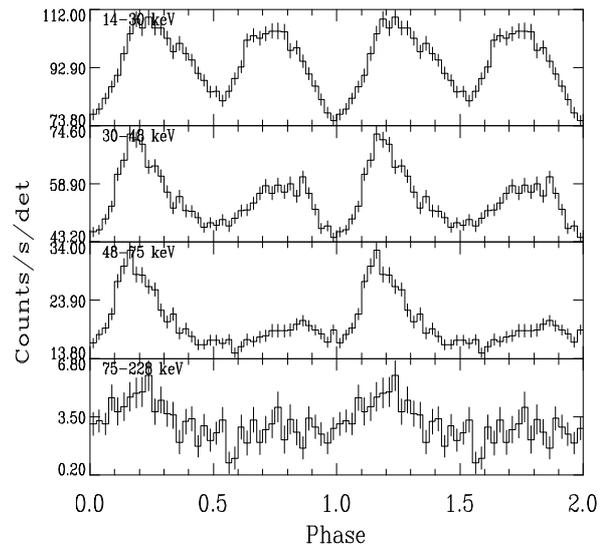

February 1994

Fig. 1. The averaged pulse profiles of A0535+26 obtained during the four large sets of the observations. The data are corrected for collimator efficiency and difference in detector efficiencies. The background has been subtracted.

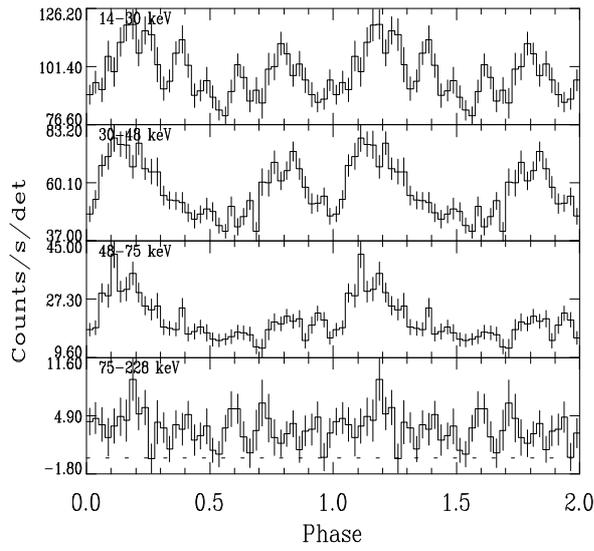

16 February 1994

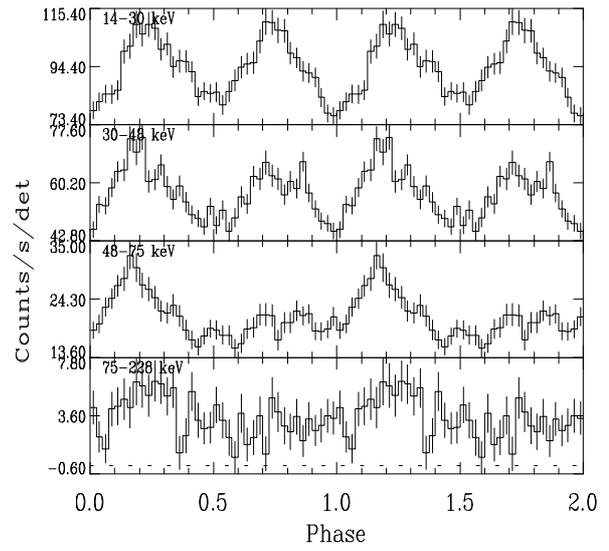

17 February 1994

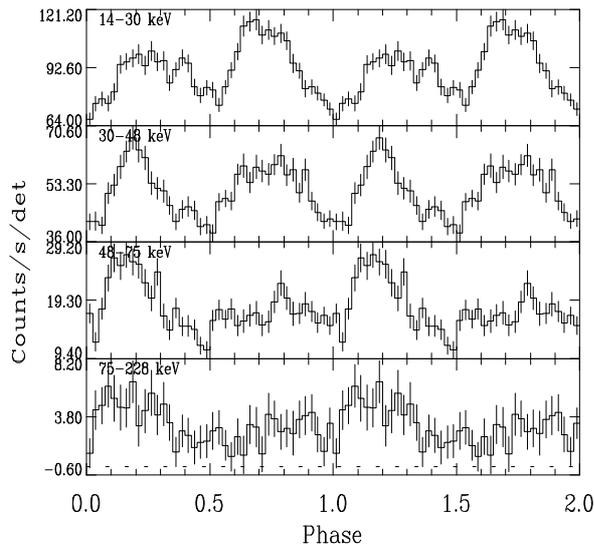

19 February 1994

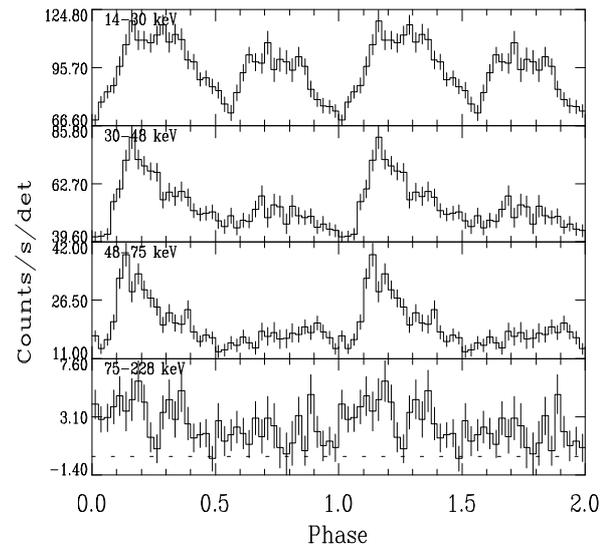

20 February 1994

Fig. 2. The averaged pulse profiles of A0535+26 obtained during the four HEXE sessions in 1994. The data are corrected for collimator efficiency and difference in detector efficiencies. The background has been subtracted.

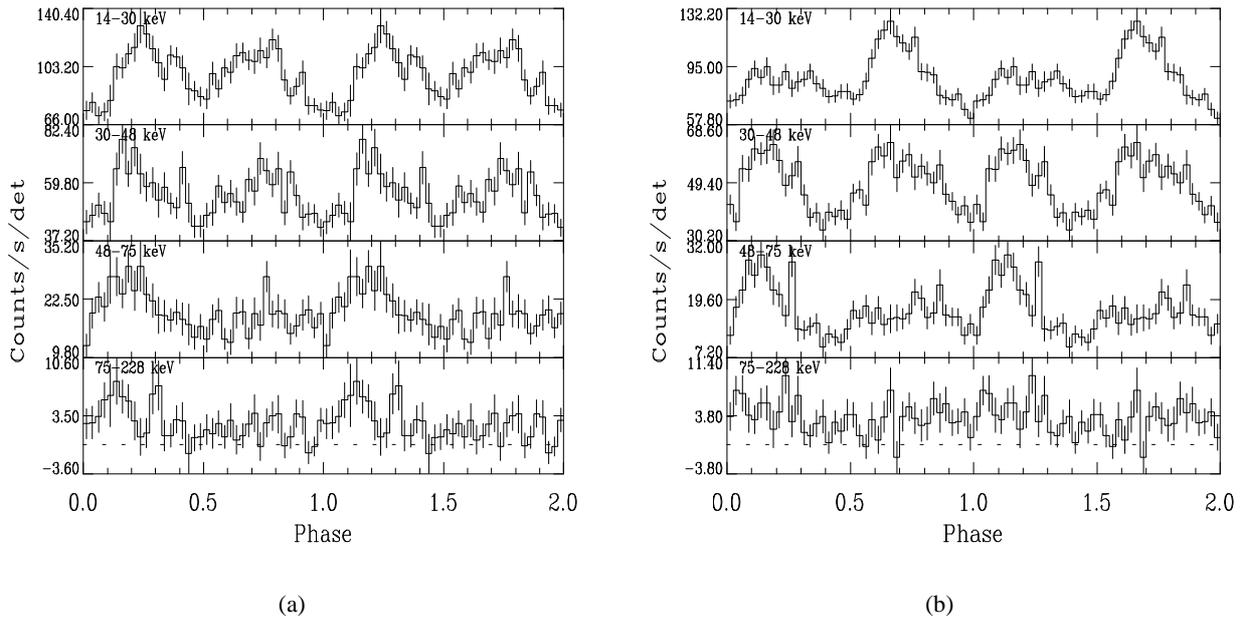

(a)                                                          (b)

Fig. 3. The averaged pulse profiles of A0535+26 obtained during the first (a) and last (b) ten minutes of the HEXE observations made on February 19, 1994. The data are corrected for collimator efficiency and difference in detector efficiencies. The background has been subtracted.

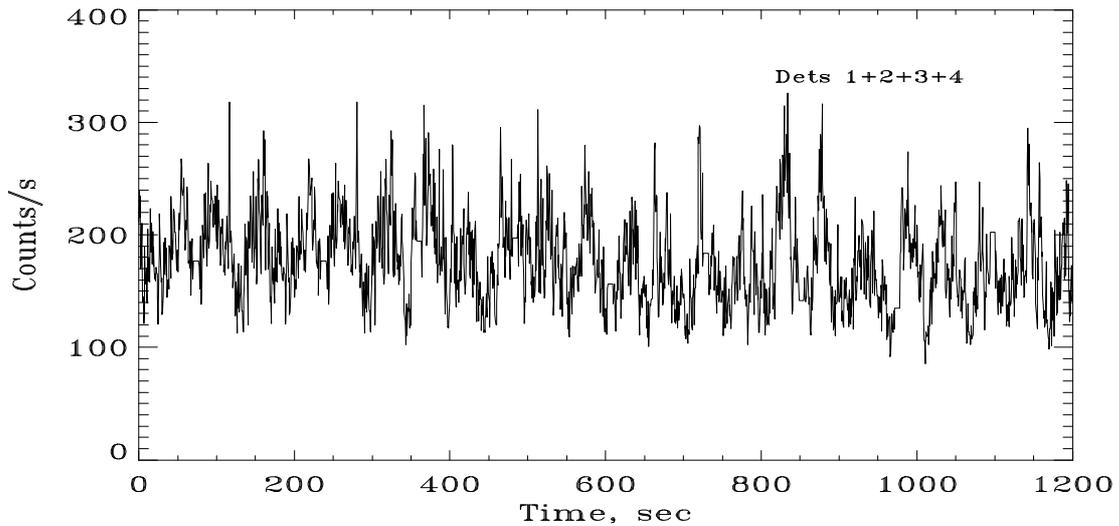

Fig. 4. The light curve of A0535+26 (the sum of the detector count rates), obtained during the observations made on February 19, 1994. The data are corrected for collimator efficiency and detector efficiency differences. The background has not been subtracted. The temporal resolution is 0.64 s.

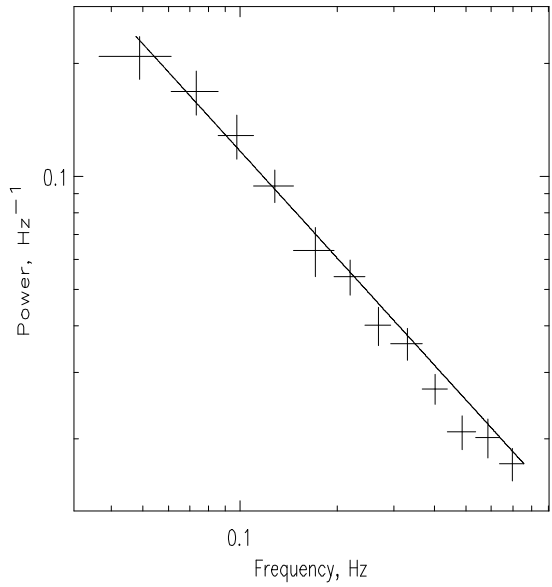

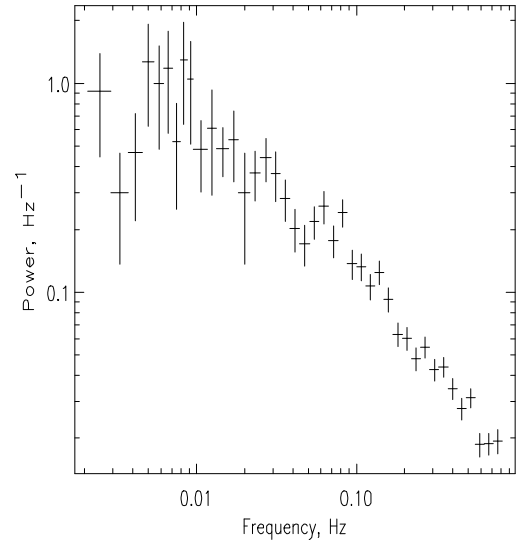

Fig. 5. The power density spectrum of A0535+26, obtained during February 1994 observations using the standard procedure.

Fig. 6. The broadband power density spectrum of A0535+26 obtained during 1994 February observations. All points, corresponding to the pulsar rotation frequency and its multiples are removed.

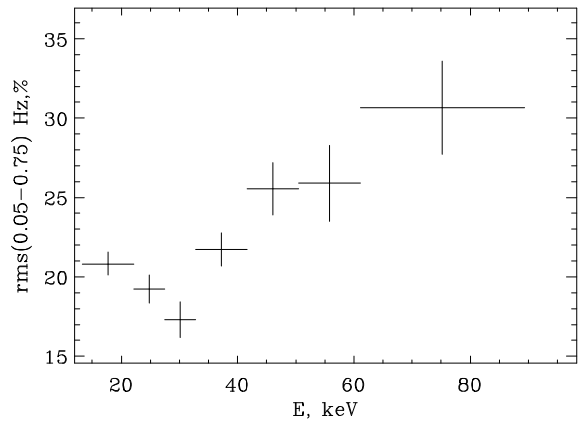

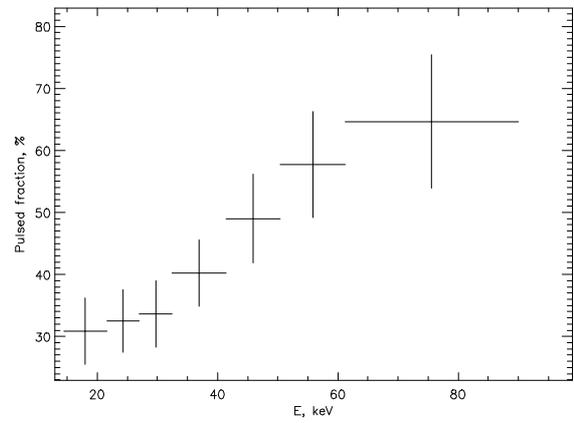

Fig. 7. The dependence of the flicker noise rms (in the frequency range 0.03-0.35 Hz) on the energy.

Fig. 8. The dependence of the source pulsed fraction on the energy for the data, obtained 1989 February 14.

Table 2.
Power density spectra parameters

| PDS | $\gamma$ | $P(f_{br})$, Hz$^{-1}$ | rms (0.05-0.75 Гц), % | $f_{br}$, Hz |
|---|---|---|---|---|
| 8/4/89 | 1.17±0.09 | 0.74±0.18 | 18.3±0.5 | 0.02 |
| 9/4/89 | 1.00±0.09 | 0.63±0.14 | 20.2±0.5 | 0.02 |
| 14/4/89 | 1.25±0.08 | 1.63±0.21 | 21.4±0.4 | 0.02 |
| xx/2/94 | 0.95±0.07 | 0.53±0.10 | 19.5±0.3 | 0.02 |
| averaged | 1±0.06 | 0.57±0.09 | 18.8±0.3 | 0.02 |
| broadband. 1994 | 1±0.06 | $0.73^{+0.42}_{-0.16}$ | 19.2±0.3 | $0.016^{+0.006}_{-0.007}$ |

In the observations of 1989 April 14 the buffered transmission mode was used. It allowed to measure the dependence of the noise rms on the energy band (Fig. 7). The noise rms grows with energy from ~20% at 20 keV to ~30% at 80 keV. For comparison, Fig. 8 shows the dependence of the pulsed fraction defined by the formula $pf = (I_{max} - I_{min})/(I_{max} + I_{min})$ on energy ($I_{max}$ is the source intensity at the pulse maximum, $I_{min}$ is the intensity at the pulse minimum). The pulsed fraction grows with energy much faster than the noise rms.

We also tried to determine whether the rms depends on the pulse phase $\varphi$. To do this we produced the distribution of the rms over the pulse phase for each session made on 1989 April 8/9 and in 1994 February, and averaged these distributions. The resultant dependence is presented in Fig. 9. It occurs that rms is consistent ($\chi^2$=23/59) with a constant value of ~23%.

The BATSE observations (Finger et al., 1996) show a peak of the quasiperiodic oscillations exist in the power density spectrum of A0536+26 at ~0.02-0.08 Hz. In our case some deviations from continuum at 0.08 Hz exist in the PDS (see Fig. 5 ), but they are not very significant, probably because of the smaller accumulated statistic.

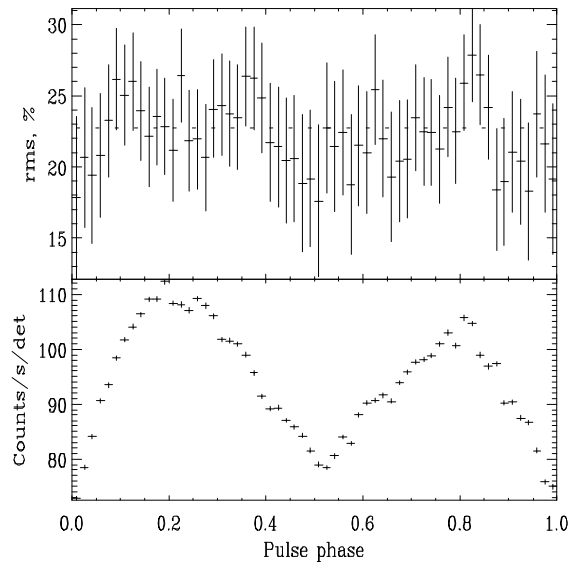

Fig. 9. The dependence of the noise rms on the pulse phase, obtained using all DTM sessions. The Poisson noise has been subtracted.

In conclusion of the discussion about the pulse profiles we mention another type of variability: strong changes of the main pulse intensity in respect to the secondary pulse intensity (see Fig. 1-3). For example, during the observations made on February 19, 1994 this peak almost disappeared in the soft energy band of the instrument, but it was still clearly seen in the harder energy channels (see Fig. 3). It is possible to distinguish the primary peak from the secondary peak by only comparing pulse profiles in the different energy bands.

## SPECTROSCOPY

An averaged source spectrum obtained during the 1994 observations is show in Fig. 10. Averaged spectra at other times all have similar shape. To investigate the source properties we approximated the spectra by the "canonical" model for X-ray pulsars (White et al., 1983):

$$I_c(E) = I_0 E^{-\gamma} \begin{cases} 1, & E < E_c \\ \exp(-(E - E_c)/E_f), & E > E_c \end{cases} \quad (1).$$

The results of the approximation are presented in Table 3. All day-averaged spectra are satisfactory described by the canonical model with a photon index $\gamma$~1.1, cutoff energy $E_c$~22.5 keV and e-folding energy $E_f$~18 keV.

Table 3. The spectra approximation by the "canonical" model for X-ray pulsars

| Spectrum | $\gamma$ | $E_c$, keV | $E_f$, keV | $\chi^2$/dof |
|---|---|---|---|---|
| 8/4/89 | 1.29±0.35 | 23.7±2.7 | 19.8±2.5 | 39/27 |
| 9/4/89 | 1.11±0.41 | 22.9±3.3 | 18.5±2.4 | 42.9/26 |
| 14/4/89 | 1.27±0.35 | 21.6±3.4 | 20.4±2.5 | 29.1/26 |
| 3/11/93 | 1.08 | 23.3±2.0 | 16.9±0.6 | 15/19 |
| xx/2/94 | 1.02±0.5 | 22.6±4.0 | 17.1±3 | 17.9/21 |
| averaged | 1.08±0.03 | 22.4±0.9 | 18.1±0.3 | 59.3/27 |
| phase 1 | 1.13±0.48 | 22.9±3.5 | 18.6±3 | 20.8/18 |
| phase 2 | 19.3 +0.3 -2.5 | 40.0 +6.6 -85.7 | 25.1 +2.0 -13.4 | 43/18 |
| phase 3 | 1.08±0.4 | 21.5±3.8 | 19±2.5 | 23.5/18 |
| phase 4 | 1.42±0.5 | 25.0±3.7 | 19.6±3 | 11.7/18 |
| phase 5 | 1.26±0.43 | 24.4±3.0 | 17.8±2.5 | 11.7/18 |

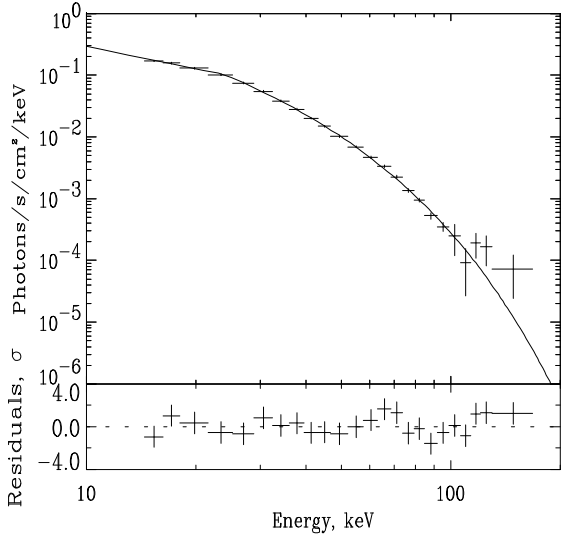

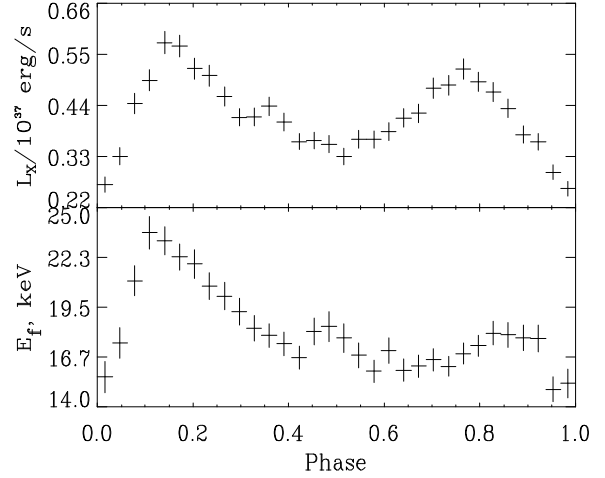

Fig. 10. The averaged spectrum of A0535+26 obtained during the observaions in February 1994.

Fig. 11. The variation of the folding energy $E_f$ in the "canonical" approximation model with pulse phase for the data of February 14, 1989

It is already clear from the pulse profile dependence on the energy band that the source spectrum varies with the pulse phase. This fact is supported by Fig. 11, which shows how $E_f$ varies with the pulse phase ($\gamma$ and $E_c$ are fixed at their average values).

For a more detailed spectroscopy it is necessary to achieve a better signal to noise ratio, so it is natural to merge adjacent phase intervals with close values of $E_f$. In the current work we used the following intervals: -0.05-0.1, 0.1-0.2, 0.2-0.4, 0.4-0.6, 0.6-0.95 (phases 1 to 5). In each of them the spectrum was approximated by the canonical model. The results are presented in Table 3 and Fig. 12. They also show the results of fitting of the spectrum averaged over all observations.

The most notable discrepancy with the model continuum is observed in the phase interval 2. Kendziorra et al. (1994) and Grove at al. (1995) interpreted this feature as a cyclotron absorption line with central energy ~100 keV. To investigate the spectrum for the possible presence of this line we multiplied the initial continuum with function

$$h(E) = \exp(\frac{-D(WE/E_1)^2}{(E-E_1)^2 + W^2})$$ (Tanaka, 1986), which reflects the shape of the cyclotron line predicted by theory.

Taking this line into account improves the fit (see Table 4). F-test indicated that the probability of a stochastic emergence of the feature in the spectrum in the phase interval 2 is ~$10^{-4}$. Note that one line at ~100 keV is sufficient to describe the data. However, including the second line at ~50 keV (see Kendziorra et al., 1994) slightly improves the fit, too. The two-line approach is especially useful for the time averaged spectrum.

In conclusion we will give the estimates of the source luminosity using the distance $d$=2.6±0.4 kpc (Janot-Pacheco et al, 1987). Extrapolating the spectrum into the energy band 2-100 keV we have (for an isotropic source): in 1989 April $L_X$~(6÷7)×$10^{37}$ ($d$/2.6 kpc)$^2$ erg/s, and $L_X$~(1÷1.5)×$10^{38}$ ($d$/2.6 kpc)$^2$ erg/s in 1993 November and 1994 February. Let us note that there are several others estimates of the distance $d$. Hutchings et al. (1978) received $d$~1.3 kpc; Giangrande et al. (1980) -- $d$=1.8±0.6 кпк.

DISCUSSION

The pulsar A0535+26 provides unique opportunity to observe different accretion regimes in the same system. During the burst an accretion rate and the source luminosity change more than 20 times (see Finger et al., 1996): from ~5×$10^{36}$ erg/s to ~1.5×$10^{38}$ erg/s. In the brightness maximum the luminosity is close to the Eddington limit.

According to the theory of accretion on a highly magnetized stars (Basko, Sunyaev, 1976) there exists a critical value of accretion rate:

$$L^* = 4 \times 10^{36} (\frac{\sigma_T}{\sigma})(\frac{l_0}{2 \times 10^5 \, cm})(\frac{10^6 \, cm}{R})(\frac{M}{M_\oplus}) \text{ erg/s}$$

($M$, $R$ are neutron star mass and radius, $M_\odot$ is mass of the Sun, $l_0$ is column length in azimuth direction, $\sigma$ is scattering cross-section, $\sigma_T$ is Thompson's scattering cross-section), at which the structure of the accretion column qualitatively changes.

When $L<L^*$ the deceleration of the accreting matter occurs in a thin layer above the NS surface with thickness ~100-100 cm. When $L>L^*$ the matter is decelerated by radiation pressure, and the accretion column height grows to tens-hundreds $R$. When $L<L^*$ most energy is radiated in the direction perpendicular to the stars surface. But when $L>L^*$ the energy is primarily radiated aside.

As it is seen from the above data, A0535+26 was observed by HEXE when its luminosity was close to the Eddington limit. In this regime the double-pike pulse profile and the strong low-frequency noise were observed.

The secondary pike strongly varies with energy, practically disappearing at $E\sim100$ keV. Comparison of parameters $E_f$ for different spectra shows that the main pulse spectrum is 1.5 time harder than the spectrum of the second pulse.

In the source spectrum at ~100 keV the significant feature is observed probably associated with the absorption at cyclotron frequency. If this interpretation is true then $B\sim10^{13}$ Gs.

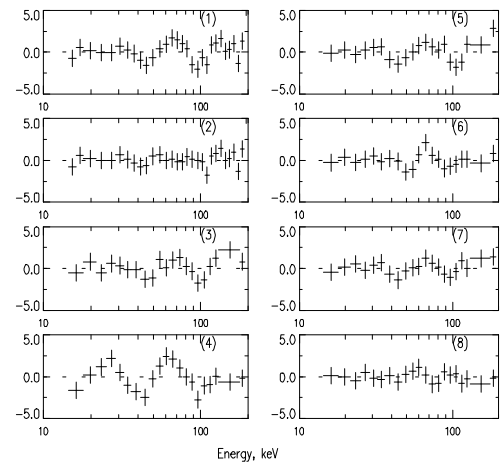

Fig. 12. The residuals (in the standard deviations) between the real and the model spectra: (1) for the spectrum averaged over all observations, (2) -- the averaged spectrum with two lines, (3)-(7) --phase intervals 1-5, (8) -- phase interval 2 with line at ~100 keV

Other methods of determinations of the magnetic field strength give

$B \approx 4 \times 10^{12}\ M_{1.4}^{3/2} L_{38}^{-3} I_{45}^{7/2} (\dot{\nu}/\ 1.2\ 10^{-11}\ \text{Hz}\ /\ \text{s})^{7/2}$ Gs for "slow" rotator (Ghosh, Lamb, 1978, 1979a,b); $B\sim2\times10^{12}$ Gs (Lipunov, 1987); $B = (1.5 - 1.8) \times 10^{13}\ \eta^{-7/4} M_{1.4}^{1/2} R_6^{-3} I_{45}^{1/2}$ Gs for the correlation between the QPO frequency and $\dot{\nu}$ (Finger et al., 1996). Here $L_{38}=L/10^{38}$ erg/s, $R_6=R/10^6$ cm, $M_{1.4}=M/1.4M_\odot$, $I_{45}$ is the moment of inertia, normalized on $10^{45}$ g cm$^2$, $\eta$ is dimensionless constant order of $0.5\div1$ (Ghosh, Lamb, 1979a,b; Ostriker, Shu, 1995).

In the strong magnetic field the scattering cross-section depends on an angle $\theta$ between the magnetic field vector and the direction of a photon propagation. This explains the variations of the spectra hardness with pulse phase.

The BATSE instrument of the CGRO observatory detected the source in the state with luminosity ~5×10$^{36}$ erg/s (Bildsten et al., 1997), i.e. when $L\leq L^*$. The pulse profiles drastically differ in the states with two different luminosities in a full agreement with the prediction by the accretion theory (Basko, Sunyaev, 1976). In particular, when $L\leq L^*$ the single pike exists in the pulse profile covering phases 0.1-0.9. This shape of the pulse profile means that the angle between the direction to the observer and axis of rotation ($\alpha$) is smaller than angle between the axis of rotation and magnetic axis ($\beta$). In addition $\beta\leq45°+\varepsilon$, where $\varepsilon$ is the polar cap half-size.

HEXE observations show that at $L>>L^*$ the pulsed fraction is ~30% (in the region far from the supposed cyclotron frequency), while at $L<L^*$ pf~90%. Simple estimates show that this relation between the pulsed fractions is hard to explain only by the variation of the visible geometric area of the cap (or column). Most probably this is the direct evidence for the amplification of pulsations due to the cross-section variations, which is predicted by various theories of accretion (Basko, Sunyaev, 1975; Meszaros, Naigel, 1985).

## CONCLUSIONS

We presented the results of the A0535+26 observations made by HEXE onboard Mir-Kvant: the power density spectrum, pulse profiles, photon spectra.

Table 4. The results of the spectrum approximation by the canonical model with lines

| Spectrum | $E_c$, keV | $E_f$, keV | $E_1$, keV | $D_1$ | $D_2$ | $\chi^2$/dof |
|---|---|---|---|---|---|---|
| phase 2 | 23.4 | 22.2±1.0 | +5.5<br>50.5<br>-3.5 | 0 | +8.3<br>3.3<br>-1.8 | 6.4/18 |
| averaged | 22.2±1.8 | 18.7±0.7 | 48.3±4.5 | 0.3±0.25 | +1.3<br>0.83<br>-0.67 | 47.5/25 |

Fixed parameters: $\gamma=1.08$, $W_1=2$ keV

The measured pulse profiles vary with energy, and the source spectrum varies with pulse phase. The analysis of this dependencies indicates that the pulsar magnetic field strength at the NS poles is ~$10^{13}$ Gs.

Comparison of the HEXE data with the data from BATSE/CGRO shows that in the high luminosity state ($L$~$10^{38}$ erg/s) the pulsar's pulse profile differs substantially from the pulse profile in the low-luminosity ($L$~$5\times10^{36}$ erg/s) state. This difference is explained by the qualitative change of the polar cap structure with formation of the accretion columns.